\documentclass[journal]{IEEEtranTIE}
\usepackage{graphicx}
\usepackage{cite}
\usepackage{multirow}
\usepackage{color}
\usepackage{amsmath}
\usepackage[font=footnotesize]{subfig}
\usepackage{float}
\usepackage{gensymb}
\usepackage{enumerate}
\usepackage{caption}

\begin{document}
\title{\huge A Reconfigurable Solar Photovoltaic Grid-Tied Inverter Architecture for Enhanced Energy Access in Backup Power Applications  }
\author{Venkatramanan~D,~\IEEEmembership{Student~Member,~IEEE}
	and~Vinod John,~\IEEEmembership{Senior~Member,~IEEE}
	\thanks{Venkatramanan~D and~Vinod John are with the Department
		of Electrical Engineering, Indian Institute of Science, Bangalore, Karnataka, India.
		e-mail: venkat86ram@gmail.com.}
	}
%
%
%
%

\maketitle
\vspace*{-3em}	
\begin{abstract}

In this paper, a photovoltaic (PV) reconfigurable grid-tied inverter (RGTI) scheme is proposed. Unlike a conventional GTI that ceases operation during a power outage, the RGTI is designed to act as a regular GTI in the on-grid mode but it is reconfigured to function as a DC-DC charge-controller that continues operation during a grid outage. During this period, the RGTI is tied to the battery-bank of an external UPS based backup power system to augment it with solar power. Such an operation in off-grid mode without employing communication with the UPS is challenging, as the control of RGTI must not conflict with the battery management system of the UPS. The hardware and control design aspects of this requirement are discussed in this paper. A battery emulation control scheme is proposed for the RGTI that facilitates seamless functioning of the RGTI in parallel with the physical UPS battery to reduce its discharge current. 
A system-level control scheme for overall operation and power management is presented to handle the dynamic variations in solar irradiation and UPS loads during the day, such that the battery discharge burden is minimized. The design and operation of the proposed RGTI system are independent of the external UPS and can be integrated with an UPS supplied by any manufacturer. Experimental results on a 4~kVA hardware setup validate the proposed RGTI concept, its operation and control. 

\end{abstract}


\markboth{IEEE TRANSACTIONS ON INDUSTRIAL ELECTRONICS}%
{}

\definecolor{limegreen}{rgb}{0.2, 0.8, 0.2}
\definecolor{forestgreen}{rgb}{0.13, 0.55, 0.13}
\definecolor{greenhtml}{rgb}{0.0, 0.5, 0.0}

\vspace*{-1em}

\section{Introduction}
Grid integration of renewable energy (RE) based distributed energy resources (DERs) has gained significant importance lately owing to the ever-increasing energy demand~\cite{Amamra, Chandra1}. A power-electronic interface is invariably required for this purpose and such a pulse-width modulated (PWM) converter is referred to as the grid-tied inverter (GTI)~\cite{Som1,blaab2}. Solar Photovoltaics (PV) based GTI systems have garnered significant attention owing to their relative advantages such as direct static energy conversion and ease of integration~\cite{Pou,blaab3}. In the event of a grid outage, a conventional GTI is designed to detect an island formation and disconnect from grid to cease DER operation~\cite{blaab1,Anti_Du}. This functionality is in compliance with the anti-islanding requirements laid out by IEEE 1547-2018 standard~\cite{IEEE}. Thus, the conventional GTI by design is incapable of harnessing the available solar energy during a grid outage period, and hence, the renewable energy resource along with the converter resources remains unutilized.

\begin{figure}
	\vspace{-3em}
	\centering

	\subfloat[]{\includegraphics[width = 0.475\linewidth]{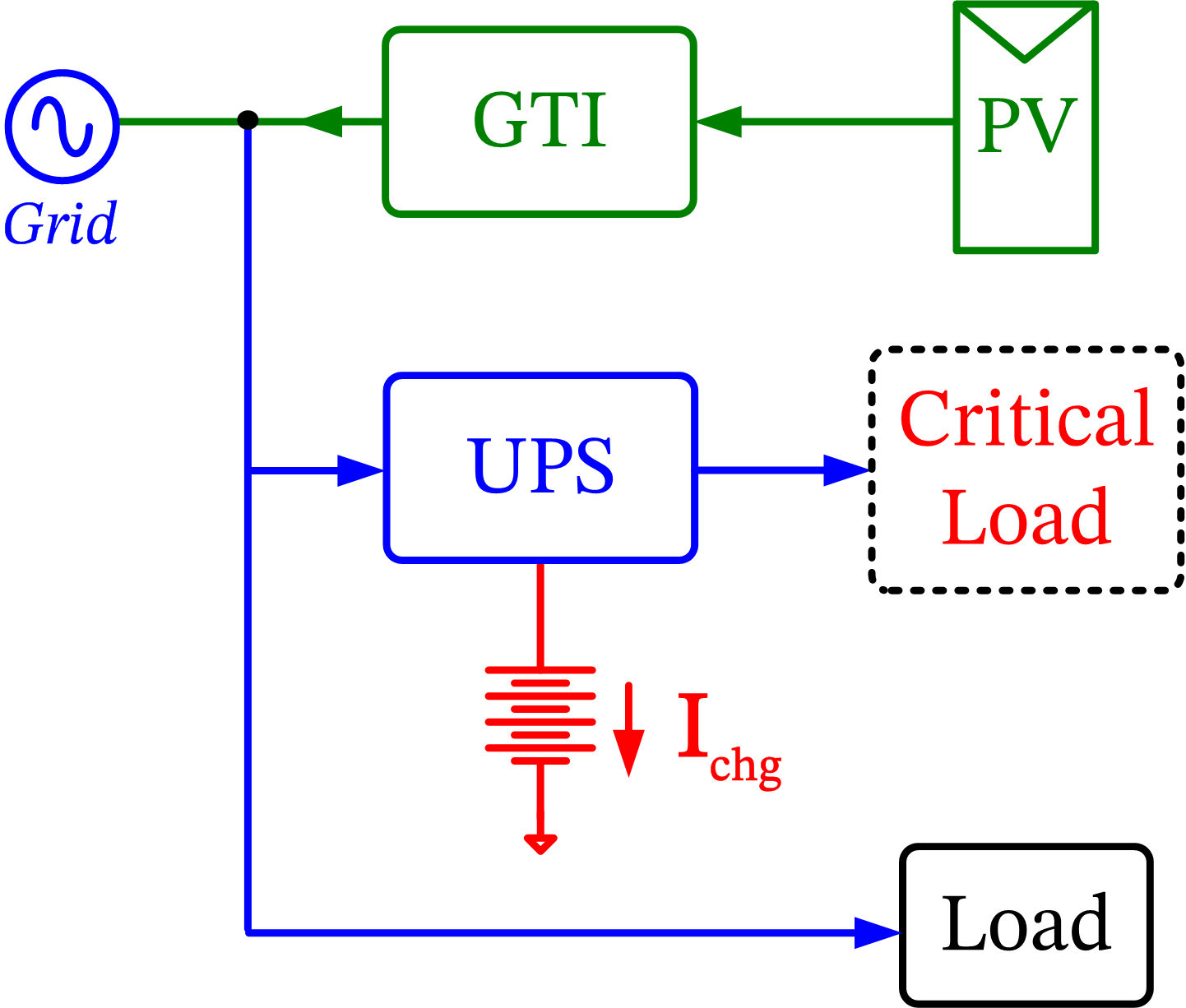}} \hfill
	\subfloat[]{\includegraphics[width = 0.475\linewidth]{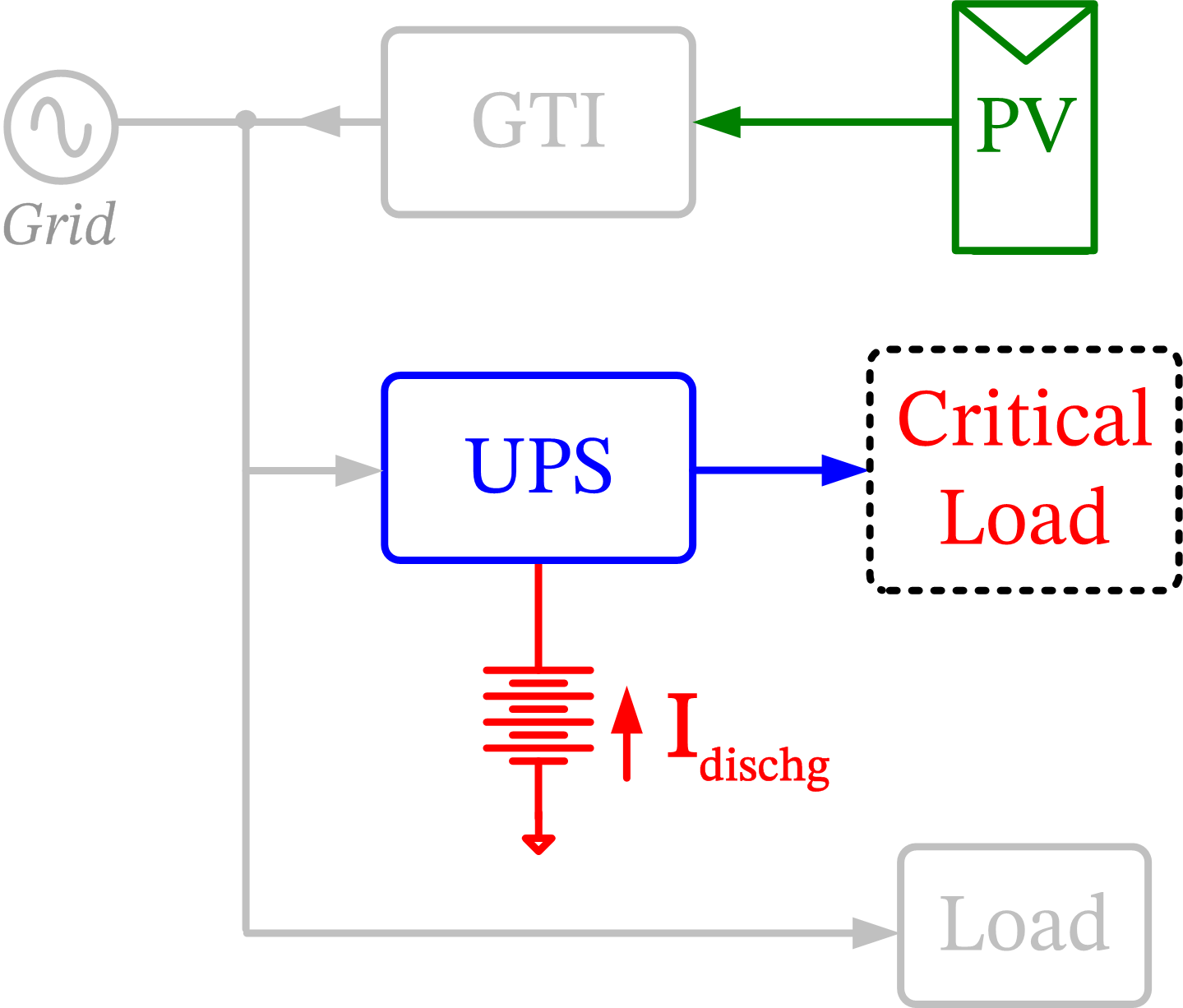}} 
	\caption{Functional block schematic of a GTI and UPS in a facility showing (a) on-grid mode and (b) off-grid mode.}
	\label{GTI}
	\vspace*{-2em}
\end{figure}

 Static uninterruptible power supply (UPS) system is conventionally employed to tackle power interruptions, which function only during grid outages to provide the necessary backup power to critical loads by discharging its battery bank~\cite{UPS2,UPS3}. It can thus be noted that while the GTI ceases to operate during a grid outage, the UPS operation for backup power is initiated only after the detection of the outage. This functional difference between the GTI and UPS is schematically shown in Fig.~\ref{GTI}, where the UPS is unable to utilize the PV energy during outage while discharging the battery to support critical loads.

\begin{figure}
	\centering
	\subfloat[]{\includegraphics[width = 0.425\linewidth]{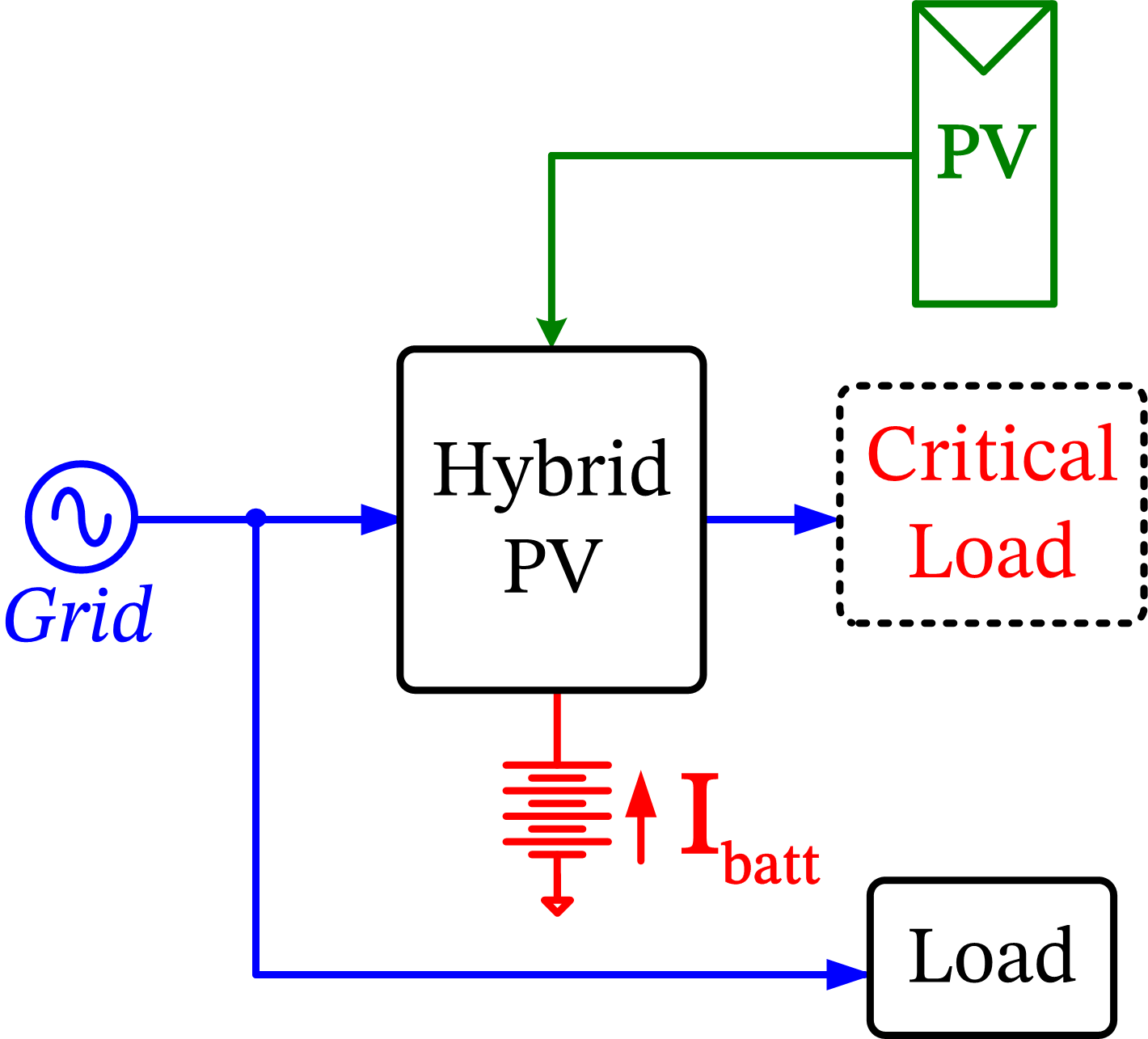}} \hfill
	\subfloat[]{\includegraphics[width = 0.5\linewidth]{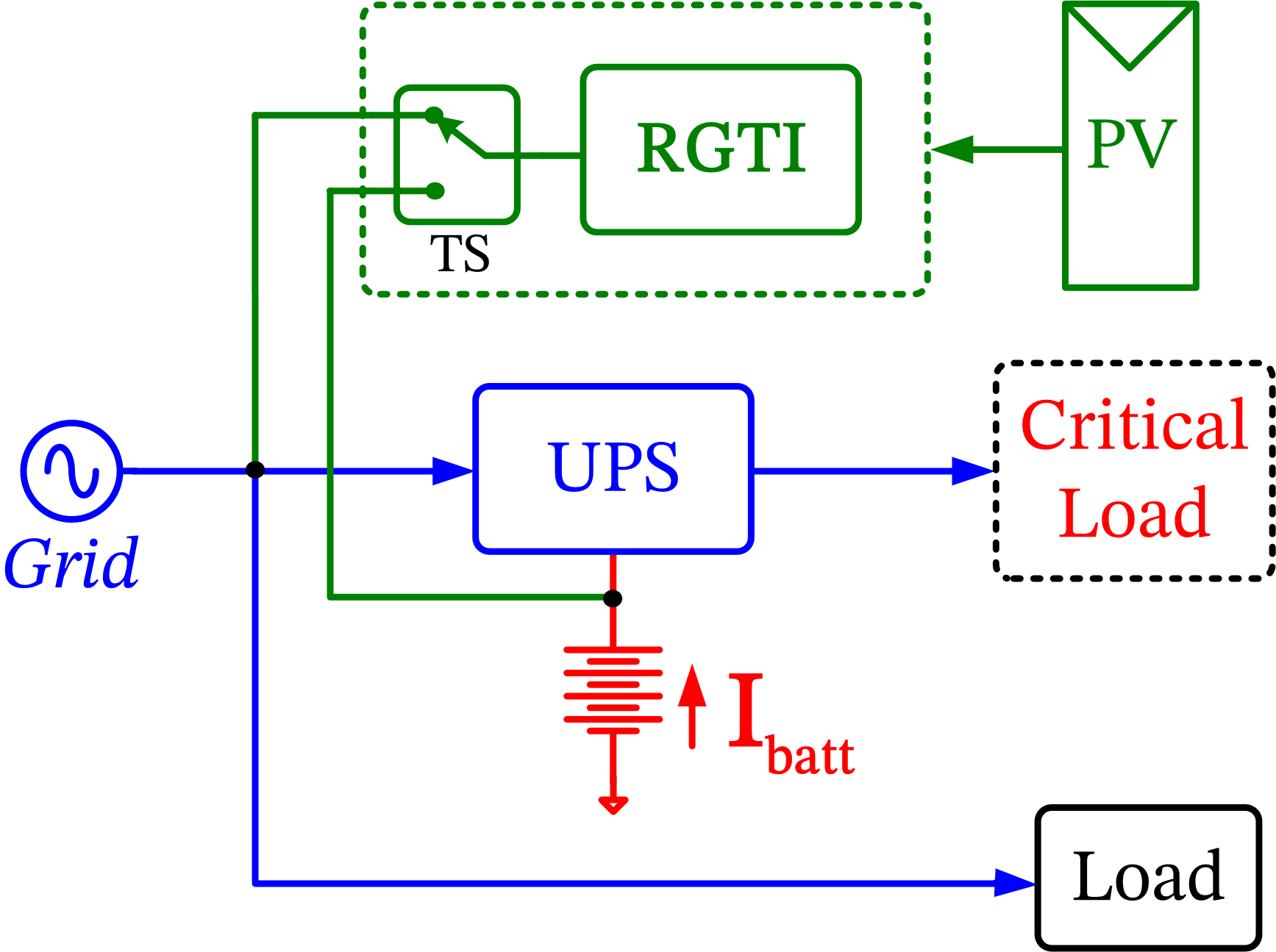}} 
	\caption{Block schematic of (a) hybrid PV system and (b) proposed RGTI system.}
	\label{DMC}
	\vspace*{-2em}
\end{figure}

In order to utilize the PV resource during a grid outage, hybrid PV architectures which combine the functionalities of GTI and UPS have been explored in the literature~\cite{arnedo2011hybrid,arnedo2011hybrid1,arnedo2011hybrid2,cavallaro2010load}. Such hybrid PV systems combine the PV with a suitable battery storage system to tackle power outages, are also known as multi-port dual-mode inverters~\cite{dual1,dual2}. These systems are capable of operating in both on-grid and off-grid modes, and the schematic of such a system is shown in Fig.~\ref{DMC}(a). In~\cite{arnedo2011hybrid,arnedo2011hybrid1}, a power circuit architecture with individual converters for PV, battery and grid is proposed, while in~\cite{arnedo2011hybrid2}, a control scheme based on instantaneous power balance theory is proposed for such a system that does not require exclusive determination of modes for operation and control. Additionally, battery management is performed along with an adaptive tolerance band based Maximum Power Point Tracking (MPPT) algorithm. In~\cite{cavallaro2010load}, a load priority based control strategy is proposed for operation and power management in a hybrid PV based UPS. 
 In~\cite{dual2}, a control strategy for a single-stage dual-mode inverter that acts as an AC microgrid is presented. In~\cite{dump2}, the hybrid PV system is extended to include a wind energy input and an active-reactive power control method is proposed for stable system operation when it is disconnected from the grid. A similar system is considered in~\cite{Hamadi}, where an improved perturb and observe P$\&$O algorithm is suggested for such a multi-input system. In many of these works, the coupling of various power conversion stages that interact with battery, PV, and grid, takes place inside the system at the common DC bus. Alternatively, there are hybrid PV systems that adopt AC-coupling technique, where the PV inverter gets tied to the AC bus at the output side~\cite{Mouli,Mouli2}. 

Although hybrid PV architectures discussed in the literature achieve the desired objective of harnessing solar energy during a power outage, they also carry the following limitations.

\begin{enumerate}
	\item The integrated systems approach with multiple energy inputs requires a unified design consideration for sizing the hardware and for the system controls. Such an integration of functionalities of the  PV converter, battery management and load side inverter into a single system leads to its increased complexity.
	\item Upgrading of an existing UPS backup system with PV capability through retrofitting is	not feasible. Many facilities have a backup power source that is previously installed. A hybrid approach requires a complete replacement of the existing backup system along with its battery-bank with a hybrid-PV system, and thus requiring a system-level change in the electrical system of the facility. 
	\item Scaling up of PV capacity without impacting or requiring a change in the balance of system is not feasible due to the integrated design approach. 
\end{enumerate}

In view of these drawbacks, a novel reconfigurable grid-tied inverter (RGTI) architecture is proposed in this work, whose block schematic is shown in Fig.~\ref{DMC}(b). In this approach, the UPS and the GTI are functionally combined while keeping the actual systems physically separate and independent in terms of design and control. The proposed RGTI acts as a regular GTI when grid is present, but reconfigures as a DC-DC charge-controller during a grid outage and interacts with the battery-bank of the external UPS. Such an operation of RGTI augments the UPS with solar power and brings PV capability to an existing backup power system. This approach is distinct from that of the hybrid-PV approach discussed in the literature. However, operation of RGTI in grid-disconnected mode with the external battery poses technical challenges when operated without communication with the UPS. It is necessary that the RGTI should not overcharge the battery or interfere with the battery management algorithms present in the UPS. Hence, suitable circuit level controls as well as a system-level supervisory control are essential for operating the RGTI in tandem with the UPS. Such an operation is the focus of this paper.

The contributions of this paper are as follows. A reconfigurable GTI scheme is proposed that enhances the capability of a conventional GTI leading to enhanced utilization of solar energy during grid outage. In order to achieve this objective, a scheme of operation of such a system is proposed in this work, where an additional battery-tied mode of operation is employed aside from the conventional grid-tied mode. In this mode, the RGTI is operated as a DC-DC charge-controller and, unlike a conventional design that always operates in MPPT mode, a new battery emulation mode is proposed for the RGTI to support the external UPS battery with solar energy. A battery-current control scheme is proposed for emulating a high ampacity battery using a PV source such that the critical load is supplied by the PV source when the insolation is adequate. Such an operation of the RGTI in tandem with the UPS ensures that the discharge burden on the physical battery is minimized, thus leading to enhanced cycle life and the overall backup system life. The circuit-level control structures for battery emulation and MPPT are presented. In order to tackle the dynamic variations in the UPS load and ambient solar irradiation and to decide on the system operating mode, a system-level control scheme is presented that functions without any communication between the RGTI and UPS. This ensures that the UPS operation and the battery management system (BMS) present in it for battery maintenance are not affected by the RGTI. Experimental results on a 5kVA hardware setup validate the operation and control of the RGTI in on-grid as well as off-grid mode. The proposed system architecture without communication allows independent design and control of the RGTI, and thus provides retrofitting capability to  operate the system in tandem with the UPS supplied by any manufacturer, which is a highly desirable feature. 
Such a solution addresses the drawbacks present in the existing hybrid-PV architectures and permits enhanced solar energy access, resource utilization, flexible PV scaling as well as extension of UPS-battery's cycle life.

The remainder of the paper is organized as follows. The coupling mechanisms of the GTI with the UPS are evaluated in Section~\ref{eval}. The proposed RGTI architecture is presented in Section~\ref{PropApp}. The grid-tied mode of operation is briefly described in Section~\ref{GTM}. The battery-tied mode of operation along with the proposed battery emulation control scheme is discussed in Section~\ref{BT}. A brief description of the MPPT control employed in this paper is provided in Section~\ref{MPPT}, followed by the description of the system-level control scheme for overall system operation in Section~\ref{Sup}. Experimental results on a hardware setup consisting of a 4kVA RGTI and a 5kVA commercial UPS are presented in Section~\ref{Expt}, which is followed by conclusion. The transfer-function models and controller structures are provided in the Appendix. 

\vspace{-1em}


\begin{figure}
	\centering
	\subfloat[]{\includegraphics[width = 0.85\linewidth]{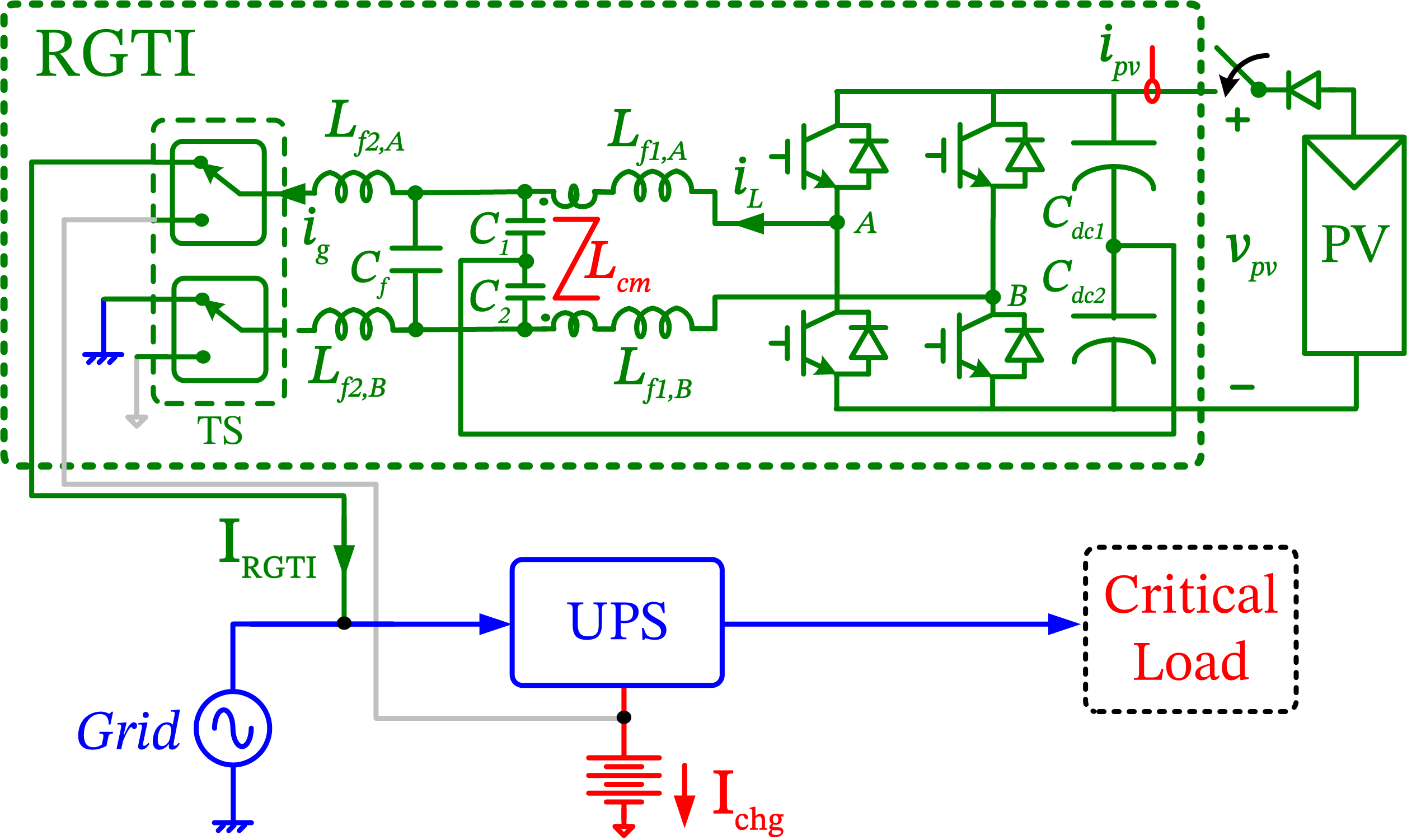}} \vspace{-1em}
	\subfloat[]{\includegraphics[width = 0.85\linewidth]{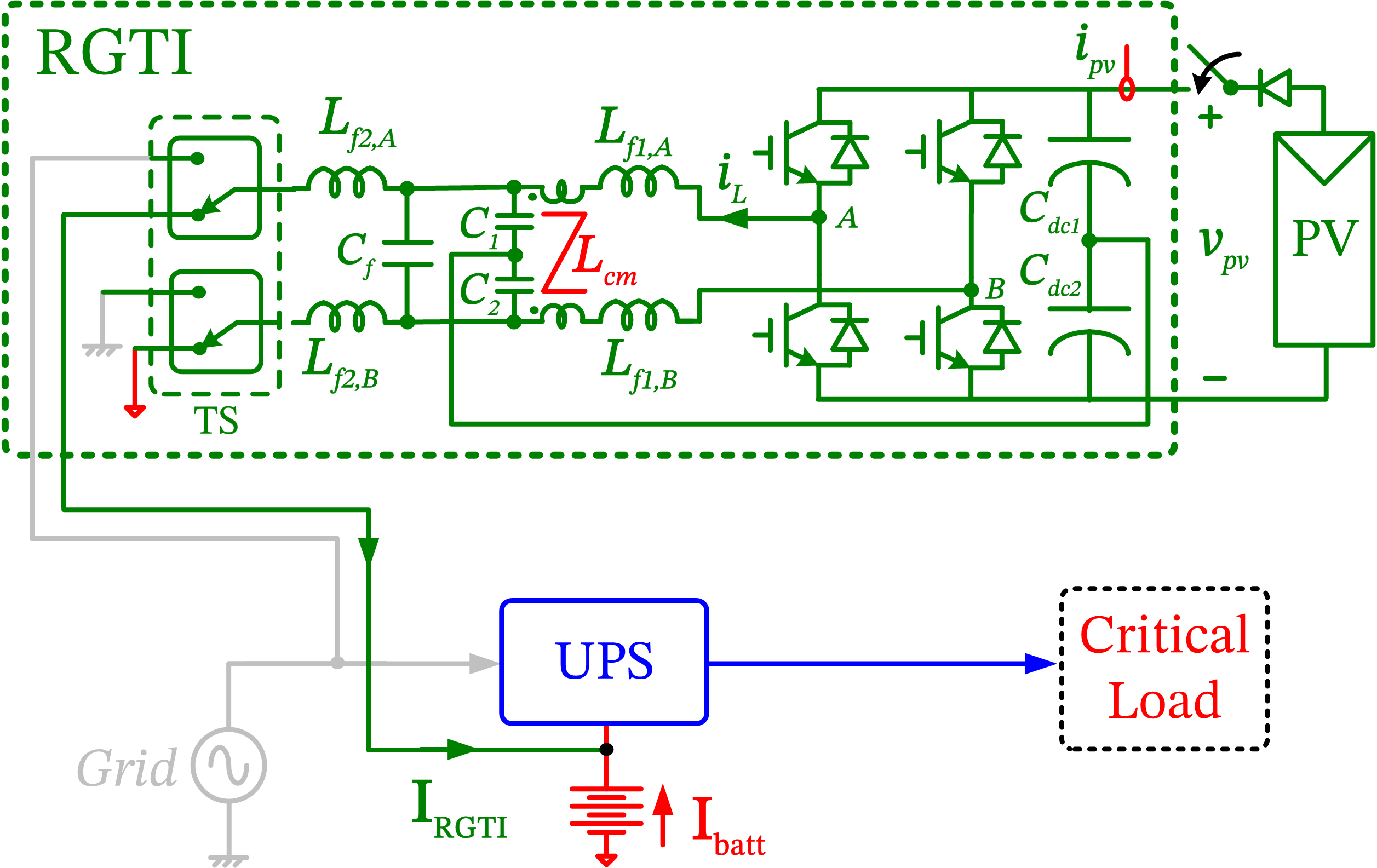}} 
	\caption{Circuit schematic of the proposed RGTI functional scheme showing (a) on-grid mode and (b) off-grid mode.}
	\label{RGTI}
	\vspace*{-2em}
\end{figure}


\section{Evaluation of Coupling Mechanisms}\label{eval}
Standard operation of a PV GTI system is as shown in Fig.~\ref{GTI}. It is intended that the GTI is not kept idle and continues to be operated to harvest solar energy during a grid outage. Hence, the GTI is coupled and operated in tandem with an external UPS. In this section, the mechanisms for coupling the GTI with the external UPS are compared and the proposed scheme for the RGTI is explained.
 \vspace{-1.0em}
\subsection{AC-Coupling }
In \textit{AC-coupling} scheme, the GTI is connected on the AC side of the UPS to share the load. In case the PV inverter is operated in voltage control mode, the UPS should be compatible for power converter paralleling in a decentralized manner. However, all UPS types may not support such a paralleling feature. On the other hand, instead of voltage control mode, if the GTI treats the UPS output as a virtual grid and interacts with the same in current control mode, then the excess power generated by the inverter during MPPT must be absorbed by the UPS to charge the battery under light UPS loading conditions. This would not only require bidirectional power flow capability in the UPS power circuit, but also the necessary supporting communication and firmware controls in place to allow such an operation without tripping the UPS. Thus, AC-coupling imposes specific hardware and firmware constraints on the UPS for compatible operation. Since varied power circuit topologies exist in UPS designs~\cite{UPS2,UPS3,UPS1}, it is not possible to assume that the UPS by default would permit such bidirectional power flows and paralleling features. Such an operation of a generic UPS through AC-coupling can lead to system tripping or potential damage, and hence this is not suitable for an independent, communicationless, reconfigurable operation of the GTI.
 \vspace{-1.0em}
\subsection{DC-Coupling }
If \textit{DC-coupling} on the battery side is employed, then the GTI and the UPS systems are decoupled in terms of operation and control, as the battery behaves as a stiff voltage source. Hence, integration of the GTI and the battery can be readily done, especially at medium power levels where the UPS battery voltage is of the order of 100~V and above. This configuration alleviates the compatibility issues between the two systems, and hence independent design and operation of the GTI and UPS are feasible. However, a suitable mechanism of control is necessary for operating the GTI in off-grid mode, which is discussed in Section~\ref{PropApp}. 
 \vspace{-1em}
\section{Proposed RGTI System}\label{PropApp}
\subsection{System Configuration}
The circuit schematic of the RGTI system configuration for single-phase single-stage PV application is shown in Fig.~\ref{RGTI}. It comprises an H-bridge inverter with unipolar PWM, differential-mode LCL-filter, common-mode (CM) LC-filter and a transfer-switch TS that enables changeover from grid to battery and vice versa. The CM filter is essential to meet the leakage current requirements of a PV inverter~\cite{CM1}. Such a hardware configuration is compatible for both DC-AC inversion and DC-DC conversion, which is leveraged in the RGTI for off-grid mode operation.
 \vspace{-1.0em}
\subsection{System Operation}
The RGTI system has two operating modes, depending on the availability of grid.
\subsubsection{Grid-tied mode} When the grid is present, the RGTI is tied to the grid and controlled to feed maximum PV power, just as a regular GTI~\cite{blaab2,blaab1}. The circuit schematic in this mode is shown in Fig.~\ref{RGTI}(a).
The converter is controlled such that MPPT is performed while maintaining sinusoidal injected grid current with the desired unity power factor~\cite{Gurr1}.


\begin{figure*}
	\centering
	\subfloat[]{\includegraphics[width = 0.625\linewidth]{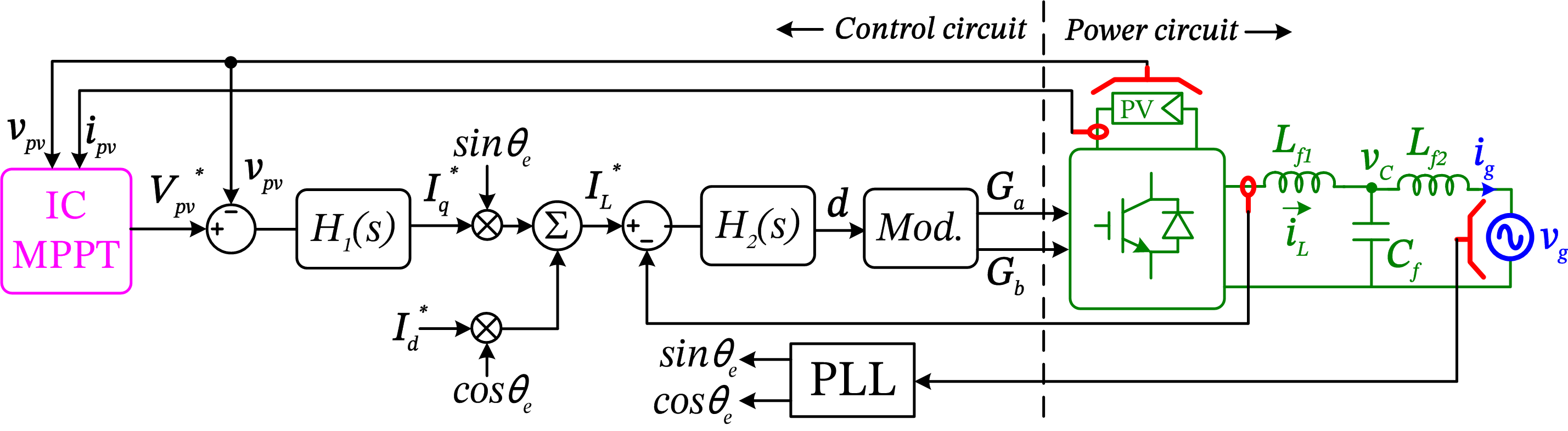}} \hfill
	\subfloat[]{\includegraphics[width = 0.35\linewidth]{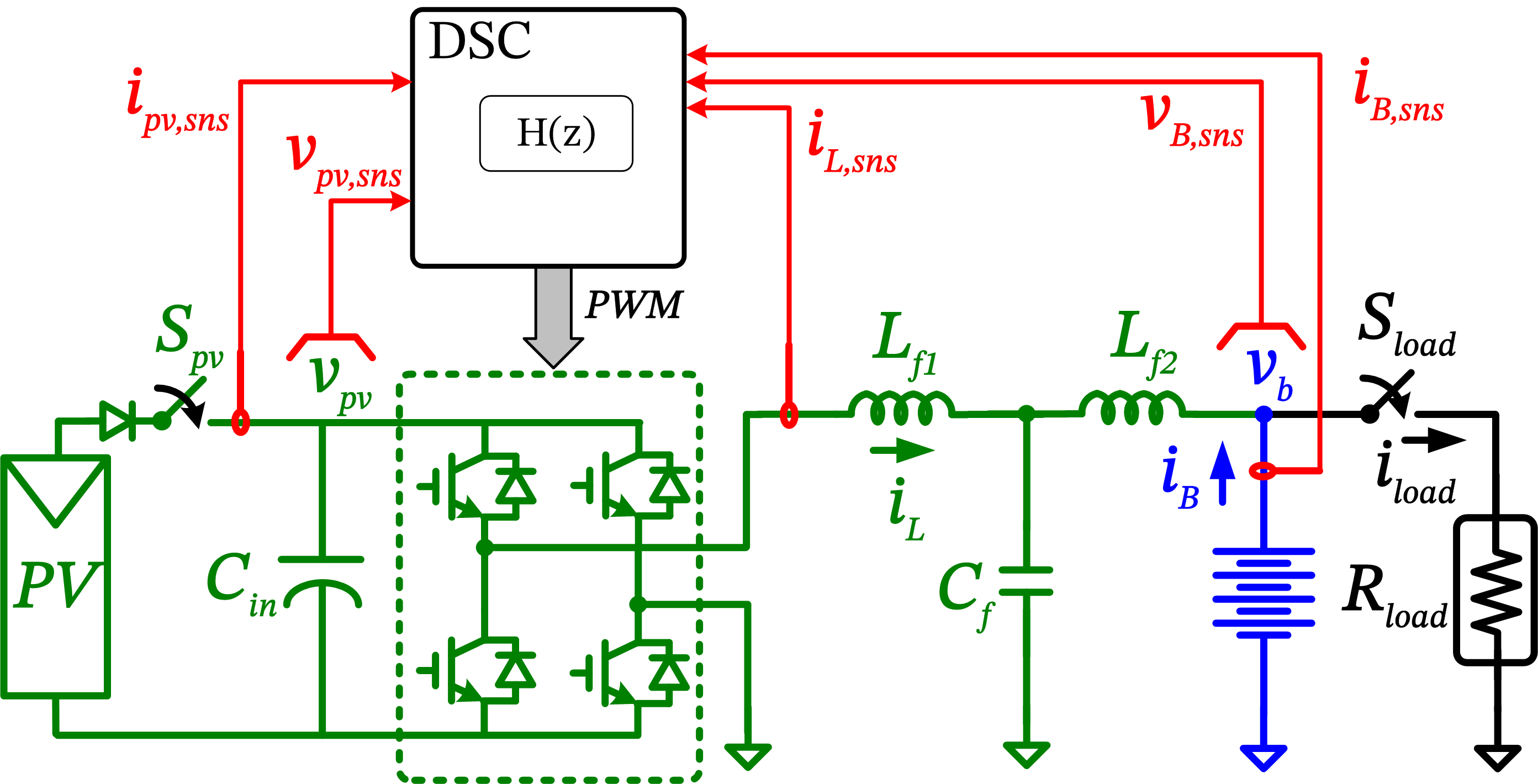}} 
	\caption{(a) Control schematic of RGTI operating in grid-tied mode performing MPPT, and (b) equivalent circuit schematic of RGTI in battery-tied mode.}
	\label{GTIcon}
	\vspace*{-1em}
\end{figure*}
\subsubsection{Battery-tied mode} In the event of a grid outage, the RGTI transitions into the battery-tied mode where it is tied to the external UPS battery and operates as a DC-DC converter. The corresponding circuit schematic is shown in Fig.~\ref{RGTI}(b). Similar to the grid-tied case, if MPPT is performed continuously by the RGTI to provide energy support, then depending on the loading conditions of the UPS and the insolation level, the battery may discharge to deliver power or charge and absorb power. This depends on the level of mismatch between generated PV power and the UPS load. Continuous MPPT operation of the RGTI can overcharge the UPS battery, leading to battery damage. However, employing the RGTI for battery management interferes with the BMS algorithm present within the UPS, and such a clash can also lead to battery mismanagement which affects the life of the storage elements. This is undesirable and, hence, a second mode of RGTI operation that complements the MPPT is necessary. This second mode is required to ensure that only adequate support is rendered by the RGTI depending on the UPS loading condition. To achieve this objective, a battery emulation mode is proposed that ensures that the discharge burden on the battery is minimized through solar power.

Thus, the RGTI is controlled in two submodes in battery-tied mode, namely, (a) MPPT and (b) battery emulation control (BEC). In MPPT mode, a voltage-based P$\&$O algorithm is employed to extract maximum PV power. In BEC mode, the RGTI is controlled in such a manner that the battery current is regulated to zero irrespective of variations in UPS load and solar irradiation. This is a critical requirement as the RGTI should not overcharge the battery during these dynamic conditions. This is discussed further in Section~\ref{BT}.

Additionally, a supervisory system-level control scheme is required to govern the transitions between these two submodes as well as the high-level grid-tied and battery-tied modes, depending on the availability of the grid, the UPS load and the ambient irradiation conditions, which is discussed in Section~\ref{Sup}. The system ratings considered and the corresponding parameters are listed in Table.~\ref{syspara}.

 \begin{table}
 	\renewcommand{\arraystretch}{1.1}
 	\centering
 	\small
 	\caption{System ratings and parameters}
 	\begin{tabular}[b]{|c|c|} \hline
 		\textbf{Item} & \textbf{Value} \\ \hline
 		Converter rating& 4 kVA \\ \hline
 		Grid voltage & 200 V \\ \hline
 		PV power rating & 3.6 kW \\ \hline
 		PV voltage $V_{pv}$ &  480 V\\ \hline
 		Battery voltage $V_{B}$  & 192 V\\ \hline
 		Battery Ampacity $C$ & 60 Ah\\ \hline
 		DC capacitor $C_{in}$, $r_{esr}$	& 	1230$\ \mu F$, 80$\ m\Omega$ \\ \hline
 		Switching frequency $f_{sw}$ & 20 kHz\\ \hline
 		Filter inductor $L_{f}$ & 2 mH\\ \hline
 		Filter capacitor $C_{f}$ & 4 $\mu F$ \\ \hline
 	\end{tabular}\\
 	\label{syspara}
 		\vspace{-1em}
 \end{table}

\vspace{-1em}
 
 \section{Grid-Tied Mode}\label{GTM}

 The control circuit schematic of the grid-tied mode is indicated in Fig.~\ref{GTIcon}(a). The H-bridge inverter is synchronized with the grid using a resonant-integrator based PLL~\cite{blaab2}. The grid voltage is aligned with the q-axis, which corresponds to the real power axis. The  inverter current is treated as the inner control variable that is controlled using a proportional-resonant (PR) current-controller $H_{2}(s)$ that tracks the AC current reference command. The structure of the current-controller and voltage-controller is designed based on~\cite{Holmes,Gurr2}.
 The selected bandwidth of current loop is 800$Hz$. An intermediate voltage loop based on a proportional-integral (PI) controller $H_{1}(s)$ is present that regulates the PV bus voltage on the DC side, the bandwidth of which is set to 5$Hz$. The voltage reference command to this loop is provided by an outer voltage-based incremental conductance (IC) MPPT controller similar to that described in Section~\ref{MPPT}. Such a three-loop control strategy for GTI is discussed in~\cite{Gurr1}.
 \vspace{-1em}
 
\section{Battery-Tied Mode}\label{BT}
The circuit schematic for the study of battery-tied mode is shown in Fig.~\ref{GTIcon}(b). For the purpose of analysis, the UPS and its output load are equivalently represented as a resistor $R_{load}$ on the battery. The split differential mode filters in Fig.~\ref{RGTI} are equivalently represented as a single-stage LC filter. 
\vspace{-1em}
\subsection{Battery Emulation Control}\label{BEC}
When two batteries are connected in parallel to feed a given load, the battery with higher ampacity takes the larger share of current. In order to emulate a battery using the RGTI, the control must ensure that the RGTI takes most of the load, such that the physical UPS battery current is minimized. Such a battery emulation scheme is feasible when the irradiation is adequate to cater to the complete UPS load, which minimizes battery discharge. This is achieved by operating the RGTI as a buck converter that regulates the battery current to zero. 
\begin{figure}
	\vspace{-1em}
	\centering
	\subfloat[]{\includegraphics[width = 0.55\linewidth]{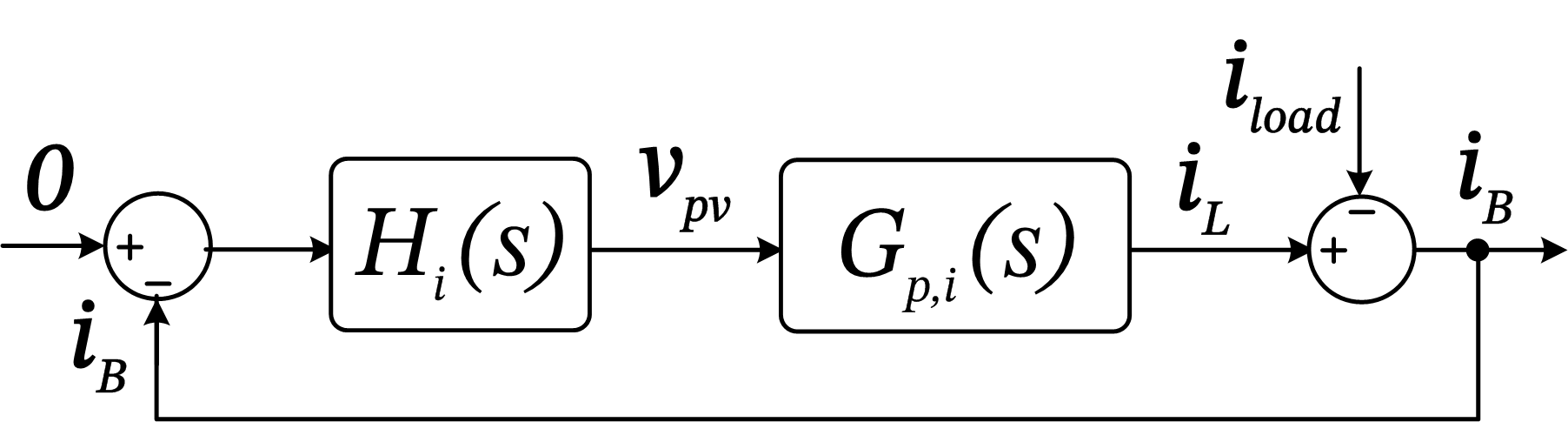}} \vspace{-2em}
	\subfloat[]{\includegraphics[width = 1\linewidth]{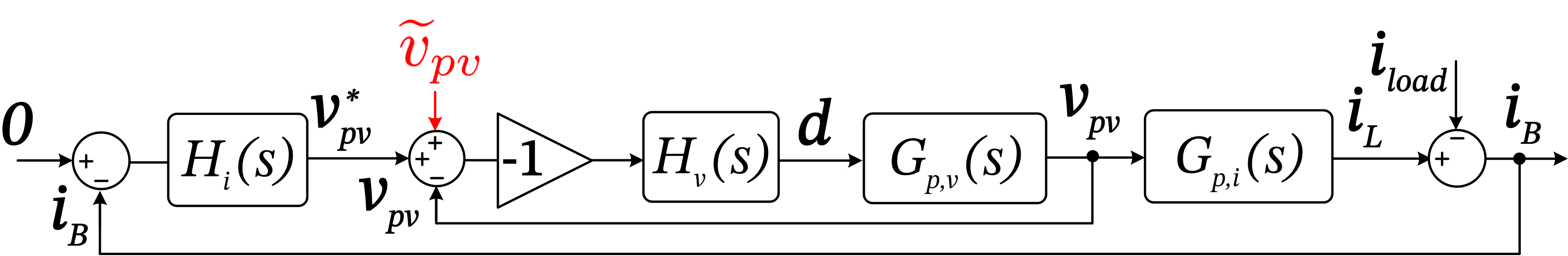}} 
	\caption{Battery emulation control schemes showing (a) basic current-control and (b) modified two-loop control.}
	\label{BE2}
	\vspace*{-2em}
\end{figure}
This is performed by treating the battery current $i_{B}$ as the feedback variable and regulating it to zero using the single-loop current-control scheme as indicated in Fig.~\ref{BE2}(a). In such a scheme, the reflected UPS load current $i_{load}$ is treated as a disturbance variable that is rejected by $H_{i}(s)$, which is a PI current-controller that achieves zero steady-state error. The system transfer-functions can thus be derived as,
\begin{gather}
\frac{i_{B}(s)}{i_{load}(s)}  = \frac{1}{1+G_{p,i}H_{i}} = \frac{s}{(1+G_{p,i}k_{p,i})s+k_{i}} \label{opnB1} 
\end{gather}
In steady-state, this yields $	I_{B} = 0 $. However, such an operation in buck mode is feasible only when the insolation is adequate to ensure that $i_{B}$ is regulated to zero. If the PV power is inadequate to match the load demand, the control loop will cause continuous decline of DC-bus voltage below under-voltage threshold, as the single-loop current-control structure operates independently with no information regarding the available PV power and the margin that exists between this value and the load demand. This will trip the buck-converter system and the power converter will cease operation. Hence, a suitable mechanism is required to detect the power margin available in the RGTI system for battery-emulation control. Instead of employing external controllable loads or irradiation sensors, the necessary detection is performed through measured electrical variables with the aid of a modified control scheme, which is explained next.
\vspace{-1em}
\subsection{Modified Control for Power Margin Detection}\label{Gdet}
It is essential to detect the change in power margin between the available peak PV power and the UPS load power, and take necessary action in order to operate the RGTI system continuously. Also, it is desirable to perform this without the use of expensive irradiation sensors. In view of this, a modified two-loop control structure is proposed, as indicated in Fig.~\ref{BE2}(b), that facilitates estimation of power margin with the aid of measured electrical circuit variables. The modified control scheme consists of an inner PV bus voltage loop and the outer battery current loop.



 \begin{figure}
 	\centering
 	\subfloat[]{\includegraphics[width = 1\linewidth]{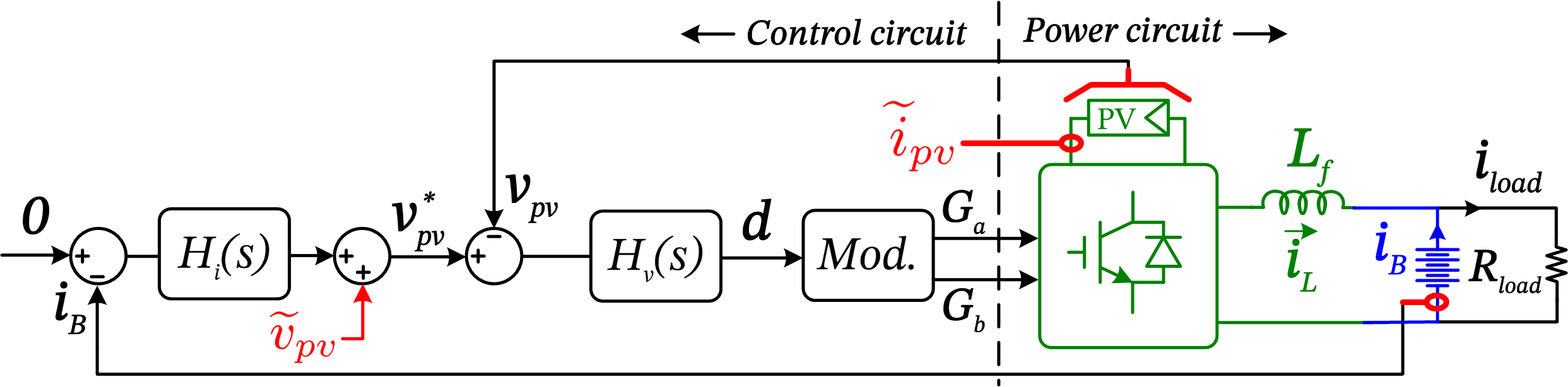}} \hfill
 	\subfloat[]{\includegraphics[width = 1\linewidth]{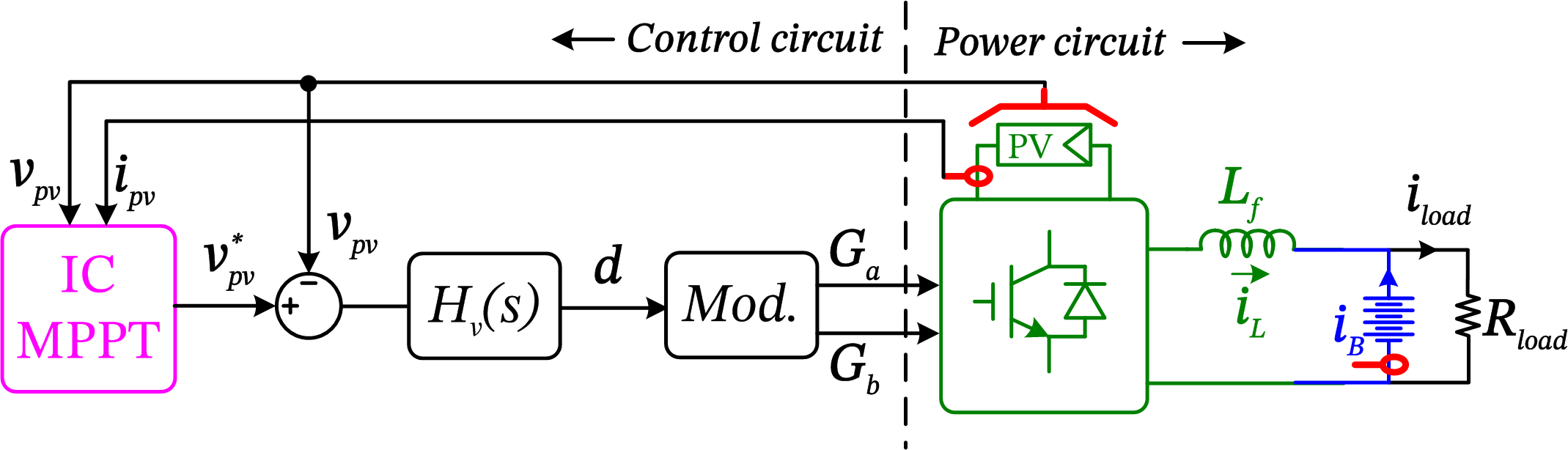}} 
 	\caption{Control schematic of RGTI in battery-tied mode showing (a) battery emulation control and (b) voltage-based MPPT control.}
 	\label{VBEcon}
 	\vspace*{-1.5em}
 \end{figure}

A perturbation $\widetilde v_{pv}$ is introduced in the inner voltage loop and the corresponding perturbation in PV current $\widetilde i_{pv}$ is measured. This is indicated in the circuit schematic of modified control shown in Fig.~\ref{VBEcon}(a). The ratio of the two quantities yields the incremental conductance of the operating point of the PV~\cite{Chap}, while catering to a given UPS load in BEC mode. The ratio of DC PV current $I_{pv}$ to voltage $V_{pv}$ in this condition yields the static conductance of the PV. The ratio of these conductances is given by,
\begin{equation}
	\widetilde{g} = \left|\frac{\widetilde{i}_{pv}}{\widetilde{v}_{pv}}\right|, \ \ \ \   G_{dc} = \frac{I_{pv}}{V_{pv}}, \ \ \ \  	G_{r} = \frac{\widetilde{g}}{G_{dc}} = \left|\frac{\widetilde{i}_{pv}}{\widetilde{v}_{pv}}\right| \frac{V_{pv}}{I_{pv}}
\end{equation}
The quantity $G_{r}$ provides an estimate of the difference between the available maximum power point (MPP) of the PV and the operating power point (OPP) for a given UPS load and irradiation condition, as explained below. A high $G_{r}$ value indicates that the PV is operating near open circuit condition, thus indicating a large margin between MPP and OPP. When OPP = MPP, $G_{r}$ is unity and indicates MPPT operation. When $G_{r}$ is close to zero., it indicates that OPP is present in the current limited region of PV close to short-circuit condition~\cite{Chap}.    
Hence, by continuously monitoring the $G_{r}$, the drift in the PV operating point towards MPP can be detected. The system-level control uses this information to handle the dynamic changes in irradiation and UPS load, and steer the RGTI system between BEC and MPPT modes, which is detailed in Section~\ref{Sup}.
\vspace{-0.5em}
\subsection{Maximum Power Point Tracking}\label{MPPT}
The control structure for MPPT operation in battery-tied mode is shown in Fig.~\ref{VBEcon}(b), where the voltage reference $v_{pv}^{*}$ is commanded by a voltage based MPPT controller based on IC algorithm~\cite{Chap}, that provides stable tracking performance~\cite{femia3}. It can be seen that the structure is similar to that indicated in Fig.~\ref{VBEcon}(a), but in this case, reference command is not provided by an outer current loop but by the MPPT algorithm. Hence, the PV bus voltage-controller design is common to both BEC and MPPT modes. 
The circuit model and the controller structure for these modes are provided in the Appendix.

   \vspace{-1em}
    \section{System-level Control }\label{Sup}

    In order to operate such an RGTI, a system-level supervisory control scheme is essential. The RGTI has grid-tied mode as well as battery-tied mode as explained in Section~\ref{GTM} and \ref{BT}, with each mode having its respective nested multi-loop circuit level control. The RGTI system operation under dynamic variations of the UPS load, solar irradiation and grid condition is explained using a mode transition table. The mode transition decisions are made using three decision variables that are continuously measured on-line by the digital signal controller (DSC). $f_{grid}$, $f_{chg}$, and $f_{G}$ are flags that indicate the status of grid, battery charging condition, and conductance ratio $G_{r}$, respectively. The status of these variables are indicated in Table.~\ref{status_RGTI}, and the corresponding mode-transition table for the system is shown in Table.~\ref{STT}. 
            \begin{table}
            	\renewcommand{\arraystretch}{1.1}
            	\centering
            	\small
            	\caption{Threshold conditions and status of measured decision variables for battery-tied RGTI state-transitions.}
            	\begin{tabular}[b]{|c|c|c|} \hline
            		\textbf{Flag} & \textbf{Status} & \textbf{Description} \\ \hline
            		$f_{G}$ & L & $G_{r}<4$, showing low power margin\\ \hline
            		$f_{G}$  & H & $4<G_{r}<1000$, showing high power margin \\ \hline
            		$f_{grid}$  & H & Grid present   \\ \hline
            		$f_{grid}$  & L & Grid absent due to islanding  \\ \hline
            		$f_{chg}$  & L & Battery discharges where $i_{B}> 0$ \\ \hline
            		$f_{chg}$ & H & Battery charges through MPPT where $i_{B}<0$\\ \hline
            	\end{tabular}\\
            	\label{status_RGTI}
            \end{table}
            
            \begin{table}
            	\renewcommand{\arraystretch}{1.1}
            	\vspace{-0.5em}
            	\centering
            	\small
            	\caption{Mode-transition table for RGTI system operation}
            	\begin{tabular}{|c|c|c|c|c|c|}
            		\hline
            		\multirow{2}{*}{\textbf{Condition}} & \multirow{2}{*}{$f_{grid}$} & \multirow{2}{*}{$f_{chg}$} & \multirow{2}{*}{$f_{G}$} & \multirow{2}{*}{\begin{tabular}[c]{@{}c@{}}\textbf{Present}\\ \textbf{Mode}\end{tabular}} & \multirow{2}{*}{\begin{tabular}[c]{@{}c@{}}\textbf{Next}\\ \textbf{Mode}\end{tabular}} \\
            		&   	&      &       &      &      \\ \hline
            		1 &	H      & -     & -    & GT\_MPPT   & GT\_MPPT   \\ \hline
            		2 &	L      & -     & -    & GT\_MPPT   & BT\_MPPT    \\ \hline
            		3  &	L      & L     & -    & BT\_MPPT   & BT\_MPPT   \\ \hline
            		4  &	L      & H     & -    & BT\_MPPT   & BT\_BEC    \\ \hline
            		5  &	L      & -     & H    & BT\_BEC    & BT\_BEC    \\ \hline
            		6  &	L      & -     & L    & BT\_BEC   & BT\_MPPT    \\ \hline
            	\end{tabular}\\
            	\label{STT}
            		\vspace{-1em}
            \end{table}
    It can be noted that when the grid is present, $f_{grid}$ = H, the system is in grid-tied mode and performs MPPT as indicated by GT\_MPPT in \textit{condition-1} in Table.~\ref{STT}. When an islanding occurs, $f_{grid}$ turns low and the system transitions into battery-tied mode as indicated by \textit{condition-2}. By default, the system enters MPPT mode as indicated by BT\_MPPT in  \textit{condition-3}, and the status of battery charging flag $f_{chg}$ is monitored. If the MPPT power is lower than the operating UPS power, then the battery effectively discharges and delivers power, and this is represented by $f_{chg}$ = L. When the load changes such that MPPT operation starts feeding power into the battery, $f_{chg}$ turns high, and the system transitions into BEC mode, which is indicated by \textit{condition-4}. Flag $f_{G}$ indicates the status of the conductance ratio explained in Section~\ref{Gdet} while the system operates in BEC mode indicated as BT\_BEC. $f_{G}$ = H indicates a high power margin available between OPP and MPP, and the BEC mode of operation can continue. However, due to variations in the ambient irradiation or due to UPS load changes, if the power margin decreases below a preset threshold, $f_{G}$ turns low, and the system transitions back to MPPT mode as indicated by \textit{condition-6}. By operating the RGTI in this fashion, the discharge burden on the physical UPS battery is minimized.
    \vspace{-1em}

      \section{Experimental Validation}\label{Expt}
\begin{figure*}
                      	\centering
                      	\subfloat[]{\includegraphics[height = 1.5in]{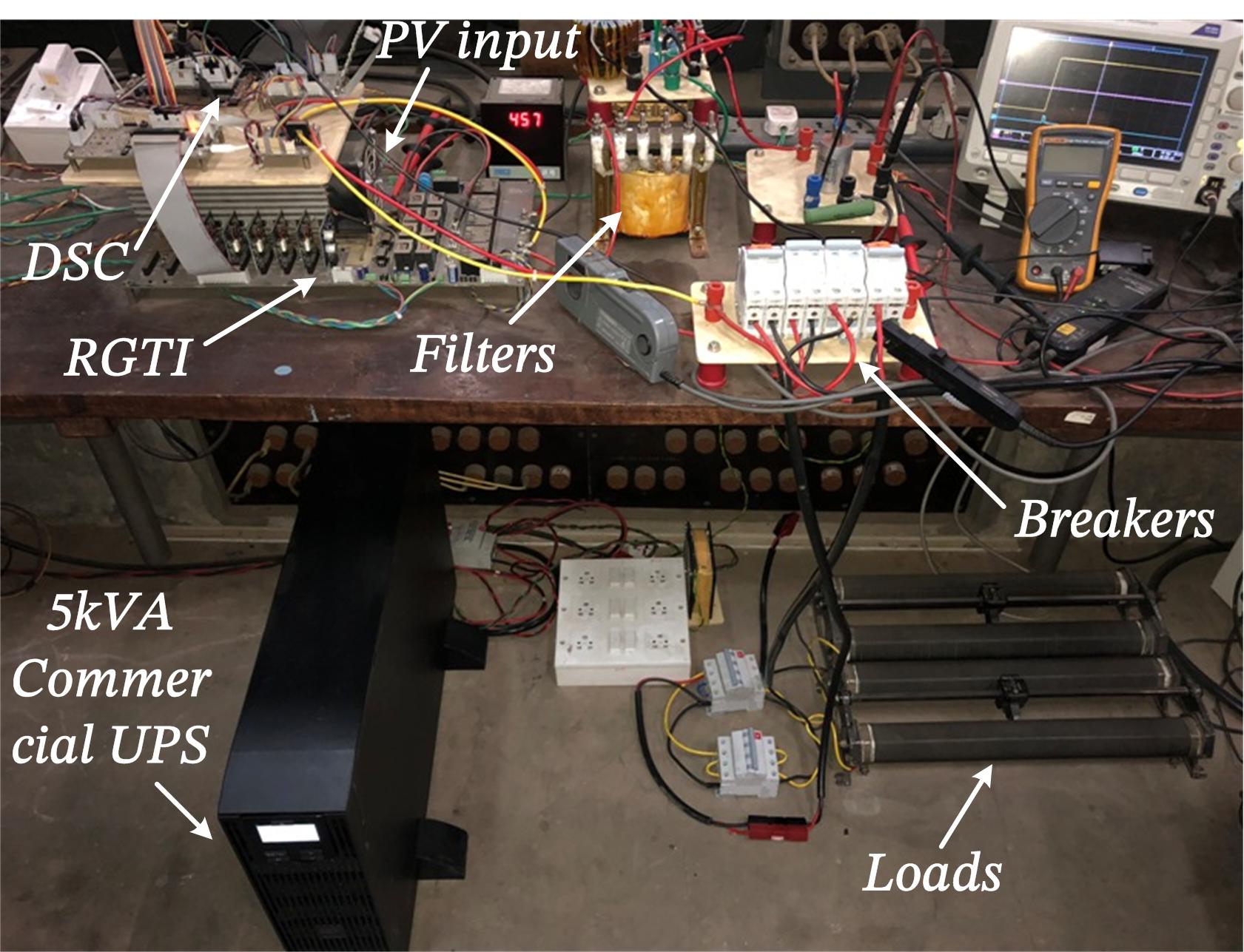}} \hfill 
                      	\subfloat[]{\includegraphics[width = 0.33\linewidth]{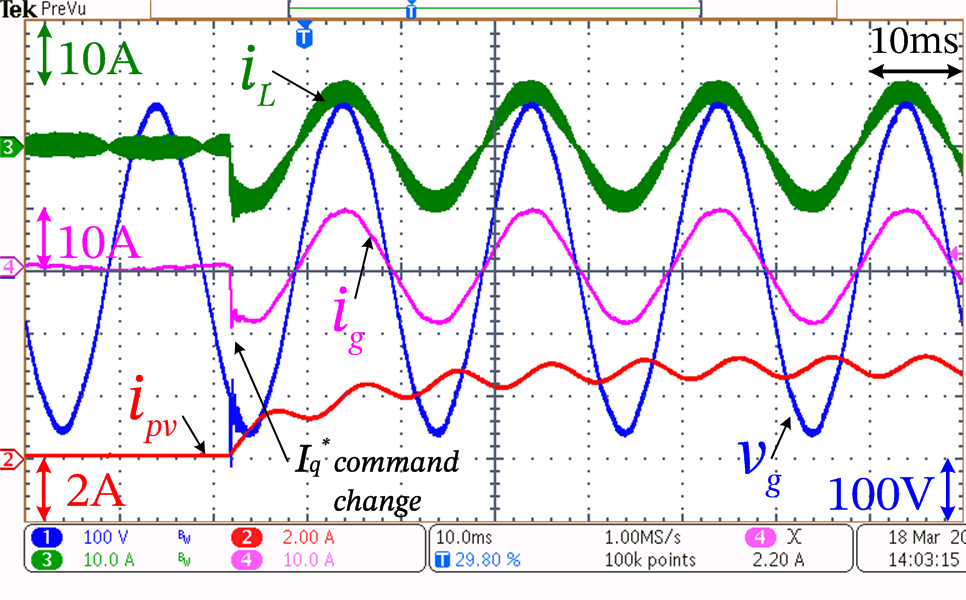}} \hfill 
                      	\subfloat[]{\includegraphics[width = 0.33\linewidth]{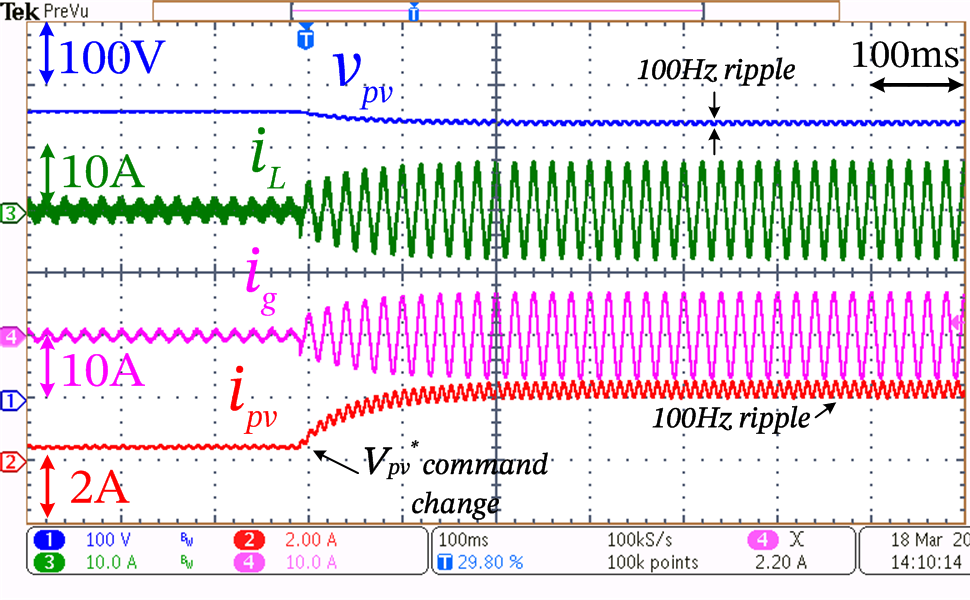}} \hfill
                      	\hfill
                      	\caption{(a) Experimental setup, and  profiles of $v_{pv}$, $i_{g}$, $i_{L}$ and $i_{pv}$ in grid-tied mode for (b) 40$\%$ step change in current loop reference $I_{q}^{*}$ for inner loop, (c) 10$\%$ step change in voltage loop reference $v_{pv}^{*}$.}
                      	\label{GT2}
                      	\vspace*{-2em}
                      \end{figure*}

      In this section, the proposed RGTI-based solution for backup power extension is experimentally validated. The hardware setup consisting of the RGTI converter, external 5kVA commercial UPS, load resistances, and contactors is indicated in Fig.~\ref{GT2}(a). The system parameters and ratings are listed in Table.~\ref{syspara}.  The system controls are implemented in a TMS320F28377S DSP based digital signal controller (DSC).
     
      Fig.~\ref{GT2}(b) shows the performance of the inner PR current-controller in grid-tied mode, for a 40$\%$ step change in real current reference $I_{q}^{*}$. As can be seen from the inverter current $i_{L}$ and injected grid current $i_{g}$, the reference is swiftly tracked by the PR current-controller well within a quarter cycle. The high performance of the PR-controller ensures a sinusoidal grid current that is devoid of switching frequency ripple due to the LCL-filter. 
      
      Fig.~\ref{GT2}(c) shows the performance of the PV bus voltage-controller in grid-tied mode, for a command change in $v_{pv}^{*}$ of 10$\%$. It can be seen that $v_{pv}$ settles in about 100ms. A corresponding increase can be seen in the amplitudes of the PV current $i_{pv}$ and the grid current $i_{g}$. 
      
\begin{figure*}
                      	\centering
                      	\subfloat[]{\includegraphics[width = 0.33\linewidth]{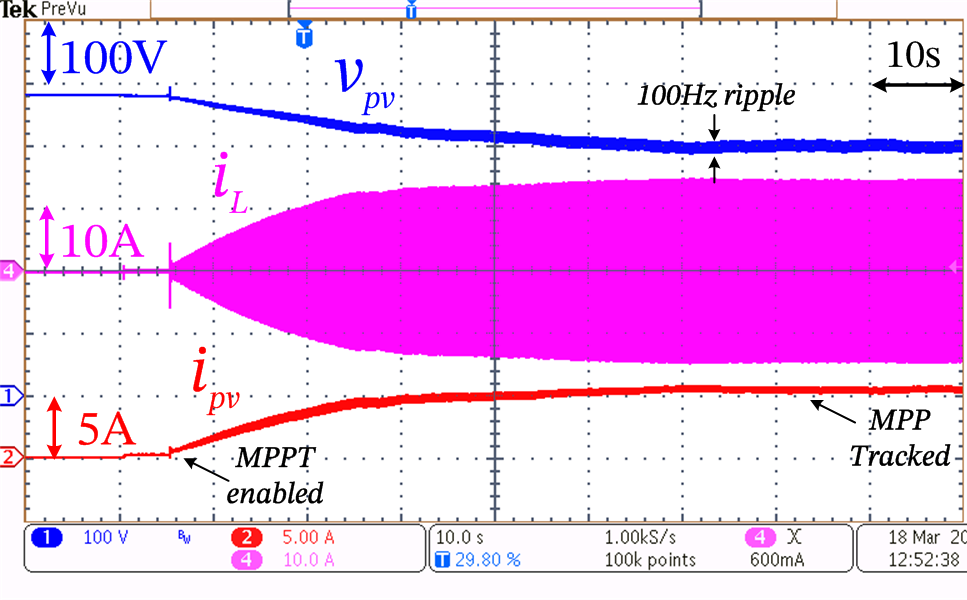}} 
                      	\hfill
                      	\subfloat[]{\includegraphics[width = 0.33\linewidth]{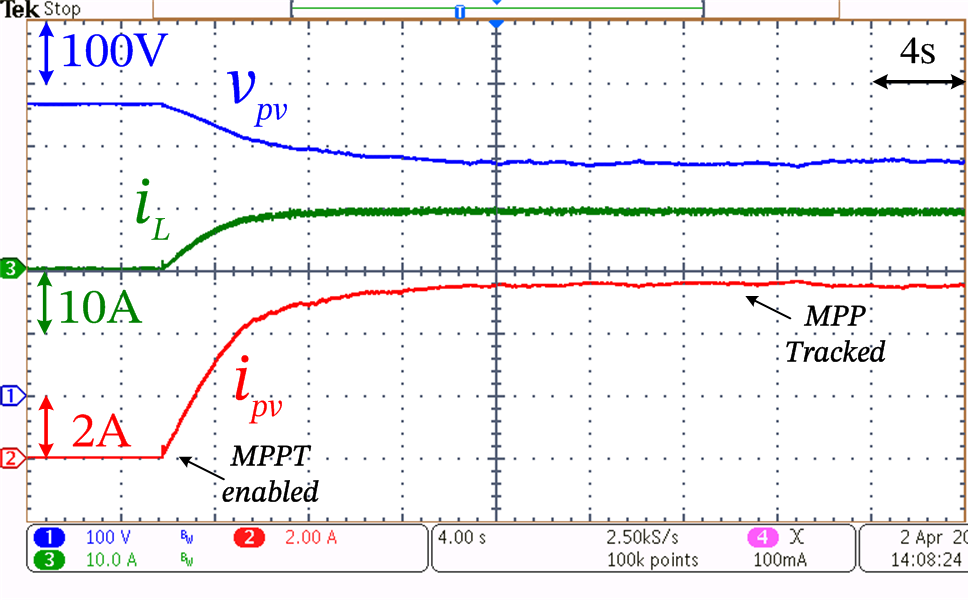}} \hfill 
                      	\subfloat[]	{\includegraphics[width = 0.33\linewidth, height =1.45in]{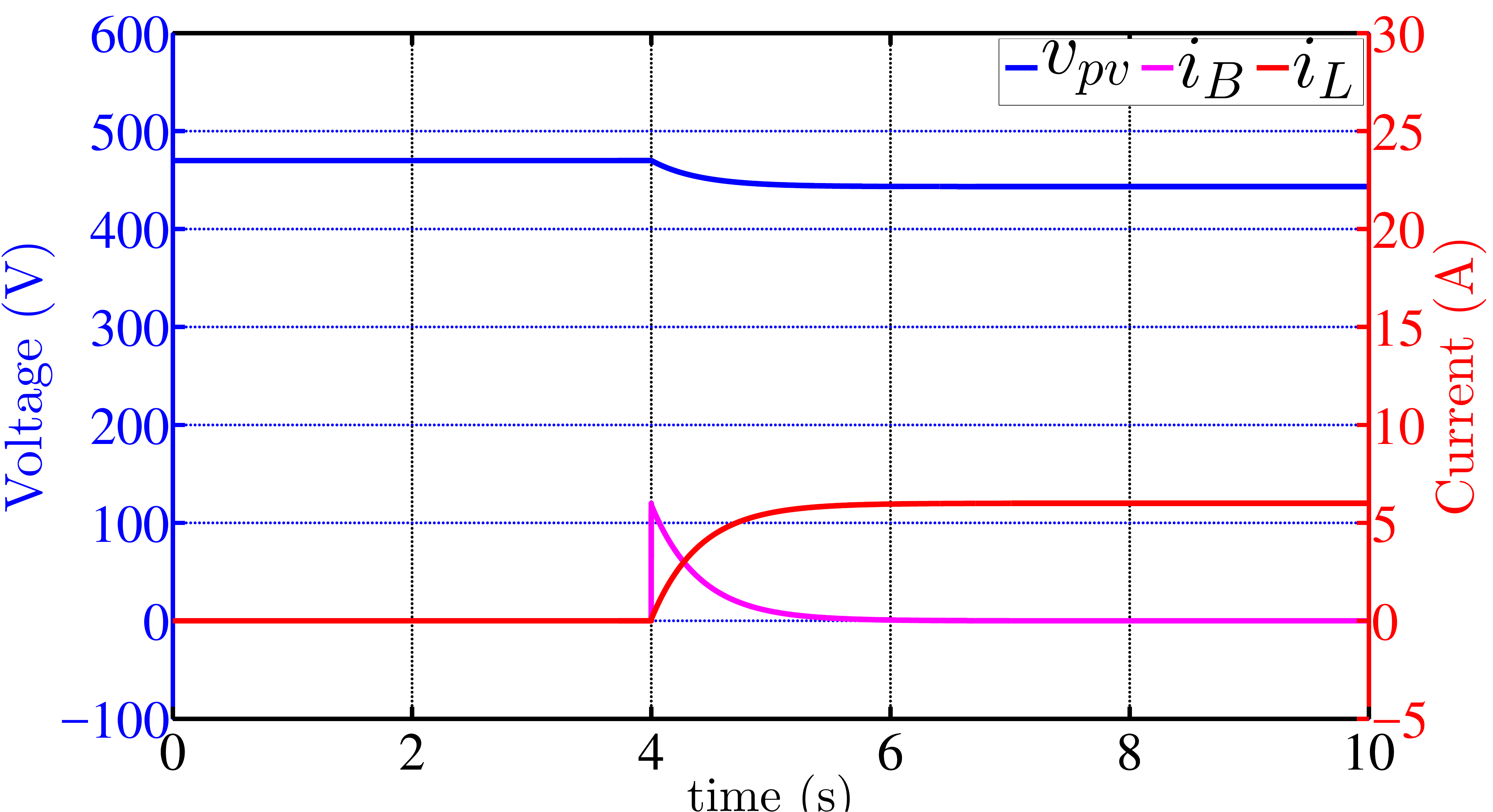}} \hfill
                      	\caption{(a) MPPT performance in grid-tied mode, (b) MPPT performance in battery-tied mode, and (c) simulated dynamic performance of battery-current controller for a 30$\%$ step change in UPS load.}
                      	\label{BT2}
                      	\vspace*{-2em}
                      \end{figure*}

      Fig.~\ref{BT2}(a) shows the MPPT performance of the RGTI in grid-tied mode, where the profiles of $v_{pv}$, $i_{pv}$ and $i_{g}$ are shown. It can be seen that after the MPPT is enabled, the amplitude of $i_{g}$ gradually increases and settles down after the tracking converges. Also, as the output power increases, the DC value of $v_{pv}$ gradually declines from OCC voltage to MPP voltage, and the amount of 100~Hz ripple seen in $v_{pv}$ and $i_{pv}$ tends to increase. This validates the capability of the RGTI to operate in grid-tied mode with MPPT.
      
         
      After detecting a grid outage, the system-level control transitions the RGTI into battery-tied mode, where MPPT is enabled. The control performance of the same is shown in Fig.~\ref{BT2}(b). The control structure of this mode is indicated in Fig.~\ref{VBEcon}(b), where the MPPT controller continuously varies the PV bus voltage reference from OCC, and the control loop tracks this reference. This process continues until the MPPT converges. Correspondingly, $v_{pv}$ is seen to reduce from open circuit condition (OCC) voltage and $i_{pv}$ is seen to increase from zero. These variables reach steady-state values indicating the convergence of MPP tracking. 
      
\begin{figure*}
              	\centering
              	\subfloat[]{\includegraphics[width = 0.33\linewidth]{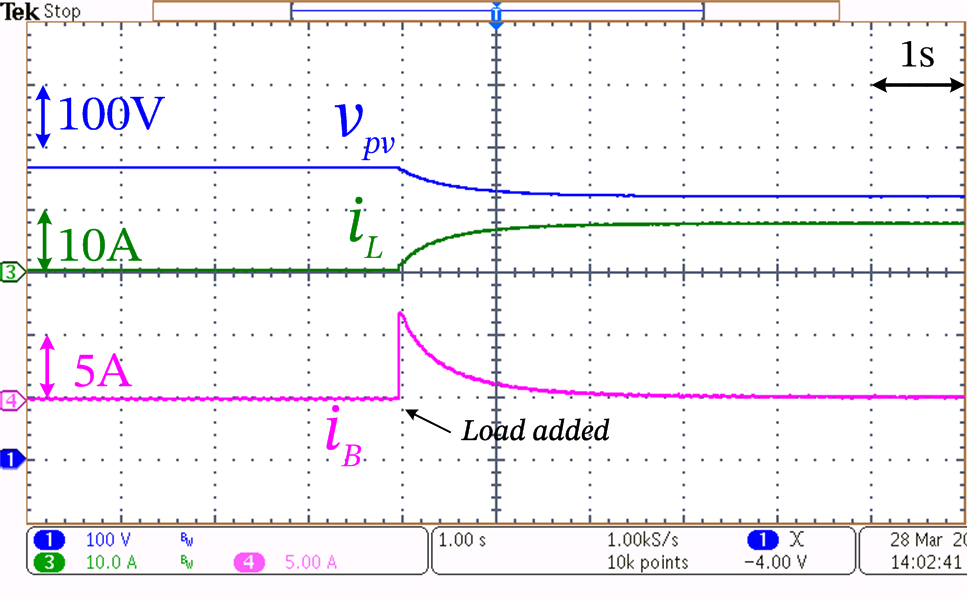}}  \hfill
              	\subfloat[]	{\includegraphics[width = 0.33\linewidth]{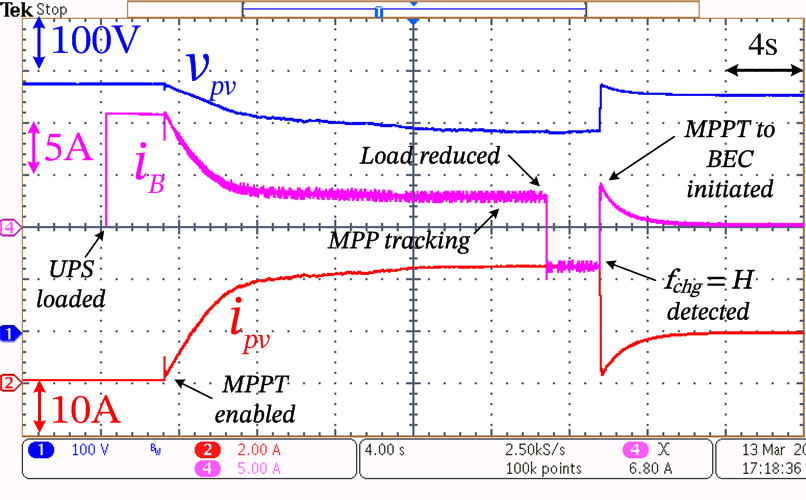}} \hfill
              	\subfloat[]{\includegraphics[width = 0.33\linewidth]{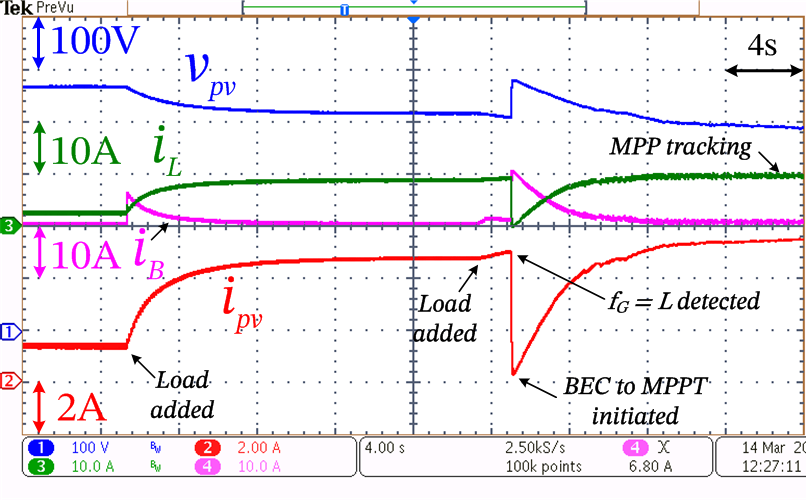}} 
              	\caption{(a) Measured performance of battery-current controller for a 30$\%$ step change in UPS load, and mode-transition performance of the RGTI during dynamic load changes showing (b) MPPT to BEC transition and (c) BEC to MPPT transition.}
              	\label{BT3}
              	\vspace*{-1.5em}
              \end{figure*}

       Fig.~\ref{BT2}(c) shows the control performance of RGTI in battery-tied mode when BEC submode is enabled,  where the dynamic responses of $i_{B}$, $i_{L}$ and $v_{pv}$ is shown from simulation based on the model discussed in Appendix, for a 30$\%$ step change in battery load. The two-loop control structure of this mode is indicated in Fig.~\ref{VBEcon}(a). The corresponding experimental dynamic response of the battery-current $i_{B}$ is shown in Fig.~\ref{BT3}(a), where a close match in settling times of the two waveforms can be noted. For a step load change, $i_{B}$ shows a step rise as the battery takes the full load initially. Subsequently the load disturbance is rejected by the control action and $i_{B}$ gradually reduces to zero, while the RGTI output inductor current $i_{L}$ increases proportionally to deliver the required power. Since RGTI functions as a DC-DC converter, $v_{pv}$ is devoid of 100Hz ripple unlike the case in grid-tied mode. This validates the capability of RGTI to operate in battery-tied mode with BEC as described in Section~\ref{BEC}, when solar irradiation is adequate.

               In Fig.~\ref{BT3}(b) a transition between MPPT and BEC submodes is illustrated. The UPS is loaded at t = 4s, and as a result the battery discharges by providing 11 A. The RGTI which is connected to the battery is enabled at around t = 8s. By default, the RGTI enters MPPT mode as explained in Section~\ref{Sup}, and correspondingly $i_{B}$ magnitude reduces. However, when the load is sufficiently high, the battery continues to discharge even when RGTI performs MPPT, as indicated in Fig.~\ref{BT3}(b). At around t = 30s, the UPS load is decreased, and this causes the battery current direction to reverse as the MPP is greater than the UPS load as explained in Section~\ref{Sup}. This condition tends to charge the battery. This is detected by the system-level control, which sets the flag $f_{chg}$ = H, and the system is subsequently transitioned into BEC mode where the battery current is regulated to zero as shown by the $i_{B}$ waveform.
           
            In Fig.~\ref{BT3}(c), a transition between BEC and MPPT submodes is illustrated. The system is initially operating in BEC mode, as indicated by the non-zero operating values of RGTI output inductor and PV currents $i_{L}$ and $i_{pv}$, respectively. At around t = 5s, the UPS load is increased and BEC operation ensures that the battery current is regulated to zero. At t = 24s, UPS load is further increased and the OPP shifts closer to the MPP of the PV, resulting in a reduction in the power margin below the set threshold. This is detected by the system-level control and the flag $f_{G}$ is set to low. Correspondingly, the system transitions into the MPPT mode. It can be noticed that the value of $i_{L}$ when MPP tracking occurs is very close to the value of $i_{L}$ in BEC mode prior to the mode transition, indicating the close proximity of OPP and MPP, which validates the detection of reduction in power margin explained in Section~\ref{Gdet}. This also verifies the system-level control and mode-transition table explained in Section~\ref{Sup}.

           \vspace{-1em}       
\section{Conclusion}
\begin{figure}
   	\centering
   	\subfloat[]{\includegraphics[width = 0.5\linewidth]{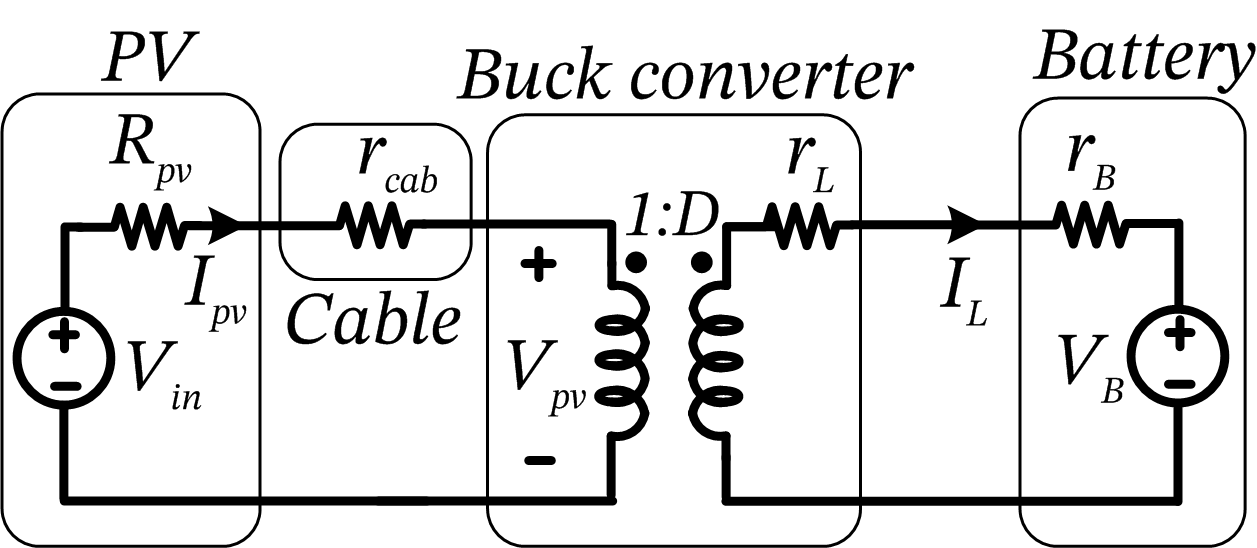}} \vspace{-1em}
   	\subfloat[]{\includegraphics[width = 0.7\linewidth]{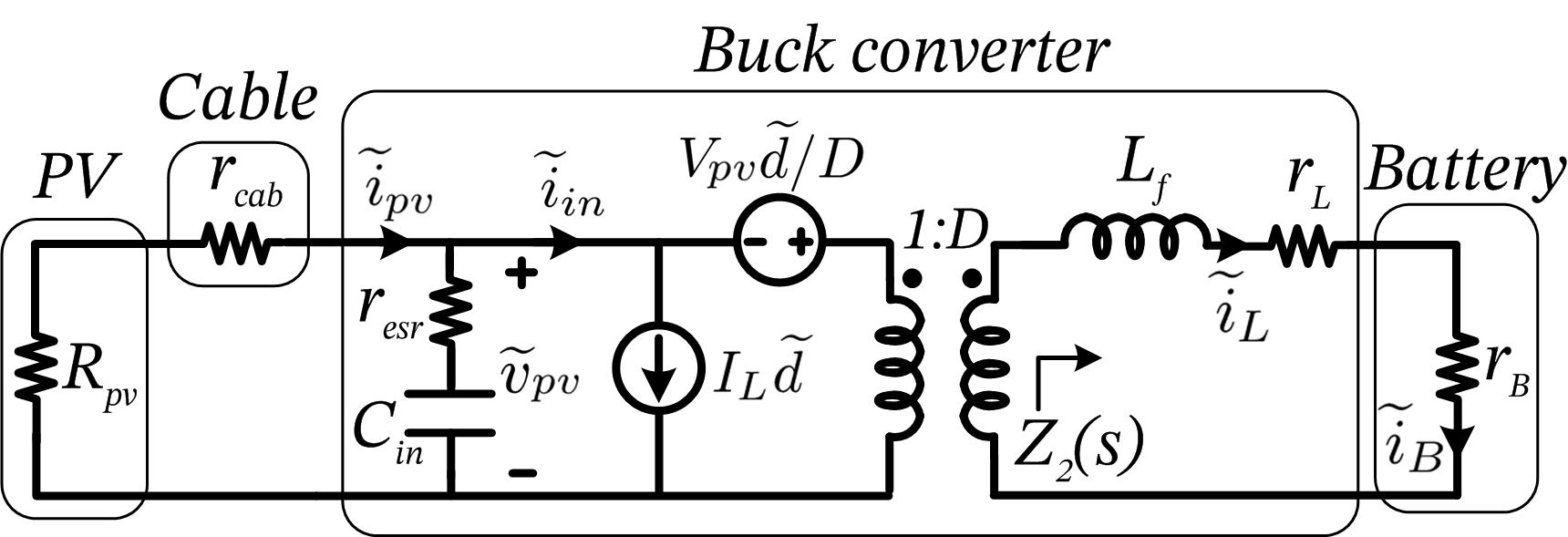}} 
   	\caption{Dynamic model of the RGTI in battery-tied mode showing (a) static DC circuit and (b) small-signal AC circuit.}
   	\label{Eqckt}
   	\vspace*{-2em}
   \end{figure}

In this paper, a reconfigurable grid-tied inverter (RGTI) architecture is proposed. This configuration departs from the conventional integrated design approach adopted for hybrid-PV systems discussed in the literature. The proposed architecture combines the functionalities of the UPS and GTI while keeping the actual systems physically separate and independent in terms of design and control. The scheme of operation of such a system is evaluated and it is shown that a DC-coupled configuration of RGTI with an UPS meets the objectives of independent design and communicationless operation. The RGTI acts as a regular GTI when the grid is present, but reconfigures as a DC-DC charge-controller during a grid outage and interacts with the battery-bank of an external UPS, thus enabling it to harness solar energy. In order to achieve this objective, a battery emulation control scheme using PV is proposed. This RGTI mode is shown to minimize the discharge burden on the battery during a grid outage whenever solar irradiation is adequate. This scheme in conjunction with the conventional MPPT mode is shown to tackle the dynamic variations in UPS load and ambient insolation conditions, while ensuring that the required energy support is provided from the PV. Discharge of the physical battery is thus reduced due to the PV support during an outage, which leads to enhanced battery operating life. A system-level supervisory control scheme is proposed that achieves such a functionality by facilitating mode transitions required for overall system operation. Experimental results on a 4~kVA hardware setup in grid-tied and battery-tied modes validate the proposed concept, circuit level and system-level control designs. Since the design and operation of the RGTI are independent of the external UPS, an existing UPS from any manufacturer can be retrofitted or upgraded with PV capability without interfering with its operation. Such a design approach thus leads to enhanced solar energy access during outage, increased resource utilization, with independent PV scaling as well as extension of UPS battery life.
\vspace{-1.0em}
\appendix 
\vspace{-0.5em}
Small-signal analysis using circuit averaging technique~\cite{Eric} yields the static DC circuit and the corresponding small-signal AC circuit are shown in Fig.~\ref{Eqckt}(a) and (b), respectively. 
 The filter capacitor $C_{f}$ of the output LCL filter causes a resonance beyond the frequency range of interest for control. Hence, its effect is ignored for control analysis, and the output filter is treated as a single L-filter $L_{f}$ = $L_{f1} +L_{f2}$. The voltage and current loop plants $G_{p,v}$ and $G_{p,i}$ are,
 \begin{equation}
 G_{p,v} = \frac{\widetilde v_{pv}(s)}{\widetilde d(s)} = k_{a}\left[\frac{(1+{s}/{{\omega_{c}}})(1+{s}/{{\omega_{d}}})}{(1+{s}/{{\omega_{a}}})(1+s/{\omega_{b}})}\right] \label{Vpvd}
 \end{equation}
  \vspace{-0.5em}
 \begin{gather}
 G_{p,i} =  {\frac{ \widetilde i_{B}(s)}{\widetilde v_{pv}(s)}} = {\frac{\widetilde i_{L}(s)}{\widetilde v_{pv}(s)}} = k_{b}\left[\frac{1+{s}/{\omega_{b}}}{(1+{s}/{\omega_{c}})(1+{s}/{\omega_{d}})}\right]  
 \label{iLVpv}
 \end{gather}
 \begin{equation}
  k_{a} {=} -R_{1}\left(\frac{2DV_{pv}-V_{B} }{D^2R_{1} + R_{2}}\right), \ 
  k_{b} {=} \frac{-1}{R_{1}}\left(\frac{V_{pv}-I_{pv}R_{1}}{2DV_{pv}-V_{B} }\right)
 \end{equation}
  \begin{equation}
 \omega_{a} = \frac{D^{2}R_{1}+R_{2}}{L_{f}}, \ \ \ \ \omega_{b} = \frac{1}{R_{1}C_{in}}  \label{p1} 
  \end{equation}
   \vspace{-1em}
 \begin{gather}
 \omega_{c} = {\omega_{a}}\left(\frac{2DV_{pv}-V_{B}}{DV_{in}-V_{B}}\right), \ \ \ \ \omega_{d} = \frac{1}{r_{esr}C_{in}} \label{z2}
 \end{gather}
For the design of inner PV voltage control loop, a PI-controller $H_{v}$(s) is designed such that the phase margin of the system is $35\degree$, with a bandwidth of $f_{bw,v}$ = $55Hz$. For the battery current regulation loop, the controller $H_{i}(s)$ employed comprises a PI-controller and a lag compensator with a corner frequency of $\omega_{n}$. The bandwidth of the current loop is set to $f_{bw,i}$ = $0.5Hz$. The small-signal perturbation $\widetilde v_{pv}$ is evaluated at $10Hz$ to obtain the incremental conductance as explained in Section~\ref{Gdet}. The current loop bandwidth is selected such that $\widetilde v_{pv}$ is not rejected.
\begin{gather}
 H_{v} = k_{p,v}\left(\frac{1+s/\omega_{c,v}}{s/\omega_{c,v}}\right) = k_{p,v} + \frac{k_{i,v}}{s}  \\
H_{i} = k_{p,i}\left(\frac{1+s/\omega_{c,i}}{s/\omega_{c,i}}\right)\left(\frac{1}{1+s/\omega_{n}}\right) \label{Hi}
\end{gather}
\vspace{-2em}
\bibliographystyle{IEEEtranTIE}
\bibliography{Manuscript_TIE}\ 
\vspace{-0.5in}
\end{document}